\documentclass[runningheads]{llncs}
\usepackage[T1]{fontenc}
\usepackage{graphicx} 
\usepackage{caption}
\usepackage{booktabs}
\usepackage{subcaption}
\usepackage{amsmath}
\usepackage{float}

\begin{document}

\title{Transcending Dimensions using Generative AI: Real-Time 3D Model Generation in Augmented Reality}

\titlerunning{Real-Time Image-to-3D Model Generation in AR}

\author{Majid Behravan\inst{1}\orcidID{0000-0001-6525-6646} \and
Maryam Haghani\inst{1}\orcidID{0009-0006-3377-0417} \and
Denis Gra{\v{c}}anin\inst{1}\orcidID{0000-0001-6831-2818}}
\authorrunning{M. Behravan et al.}
%
\institute{Virginia Tech, Blacksburg VA 24060, USA }

\maketitle

\begin{abstract}
Traditional 3D modeling requires technical expertise, specialized software, and time-intensive processes, making it inaccessible for many users. Our research aims to lower these barriers by combining generative AI and augmented reality (AR) into a cohesive system that allows users to easily generate, manipulate, and interact with 3D models in real time, directly within AR environments. Utilizing cutting-edge AI models like Shap-E, we address the complex challenges of transforming 2D images into 3D representations in AR environments. Key challenges such as object isolation, handling intricate backgrounds, and achieving seamless user interaction are tackled through advanced object detection methods, such as Mask R-CNN. Evaluation results from 35 participants reveal an overall System Usability Scale (SUS) score of 69.64, with participants who engaged with AR/VR technologies more frequently rating the system significantly higher, at 80.71. This research is particularly relevant for applications in gaming, education, and AR-based e-commerce, offering intuitive, model creation for users without specialized skills.

\end{abstract}



\section{Introduction}
\label{sec:intro}
 
The accelerated development of generative AI has significantly impacted many areas, leading to notable advancements natural language processing, image generation, and more.
One of the most exciting developments is the ability of generative AI to create three-dimensional (3D) objects.

Generative AI models have proven to be incredibly versatile, enabling the generation of 3D models from descriptive text and visual inputs.
This advancement is particularly crucial for AR, where 3D objects play a significant role in creating immersive experiences.
By leveraging generative AI, we can develop AR applications more efficiently, reducing both time and cost.
However, manually creating 3D content is a time-consuming process that requires a high level of expertise~\cite{gao2022get3d,chan2022efficient}.

The power of generative AI models lies in their ability to overcome the limitations of pre-designed objects.
Users can now create an unlimited number of 3D objects, tapping into their creativity to design innovative spaces and experiences.
This capability democratizes 3D model creation, making it accessible to a broader audience and fostering a new wave of creativity and innovation~\cite{cao2020,hanocka2019}.

The integration of generative AI with AR has the potential to revolutionize various industries, from gaming and entertainment to education and retail.
By enabling the creation of realistic and customizable 3D objects, we can enhance user experiences and open up new possibilities for interactive and immersive applications~\cite{muller2011}.

Our research focuses on the technical aspects of this integration, exploring the algorithms and methodologies required to achieve high-quality 3D model generation from diverse sources.
We delve into the challenges and solutions associated with converting images into accurate and visually appealing 3D models in AR.
The implications of our work extend beyond the realm of AR, as the techniques developed can be applied to virtual reality (VR) and other immersive technologies~\cite{nash2022,nobari2021}.
The ability to generate 3D models on-the-fly enhances the flexibility and adaptability of these technologies, making them more responsive to user needs and preferences.

In this research, our contribution is the enhanced capability of generative AI models to convert images into 3D models within AR environments.
We present an approach that allows for the seamless transformation of real-world objects and items from online stores web pages into the AR space.
This approach bridges the gap between the physical and digital worlds, providing users with a powerful tool to create and interact with 3D objects in real-time.
With tools like these, designers and developers can achieve new heights of productivity and creativity, fundamentally transforming how we interact with digital content.

\begin{figure*}[ht]
  \centering
  \begin{minipage}{0.24\textwidth}
    \includegraphics[width=\linewidth]{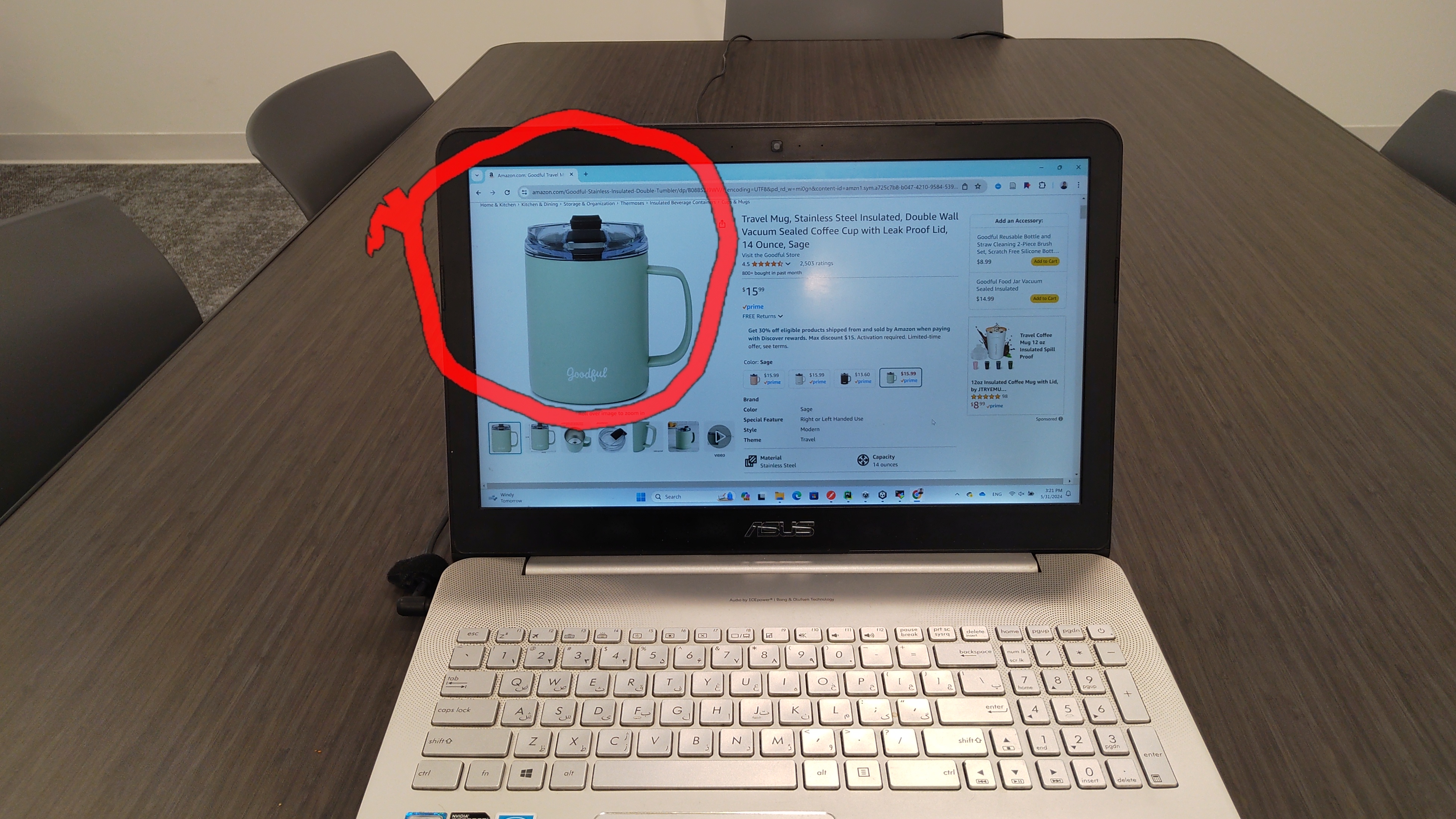}
    \label{fig:image1}
  \end{minipage}
  \begin{minipage}{0.24\textwidth}
    \includegraphics[width=\linewidth]{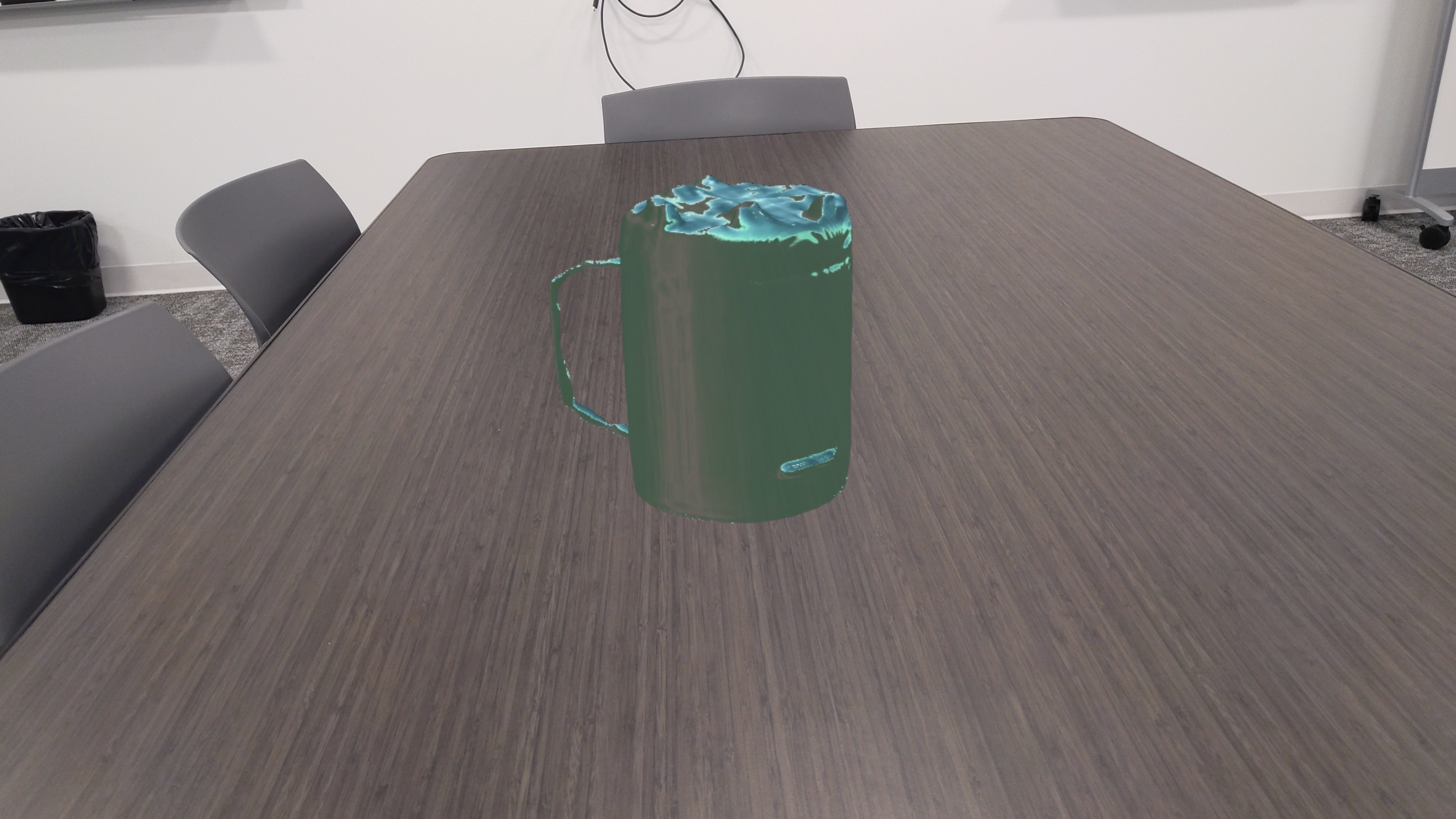}
    \label{fig:image2}
  \end{minipage}
  \begin{minipage}{0.24\textwidth}
    \includegraphics[width=\linewidth]{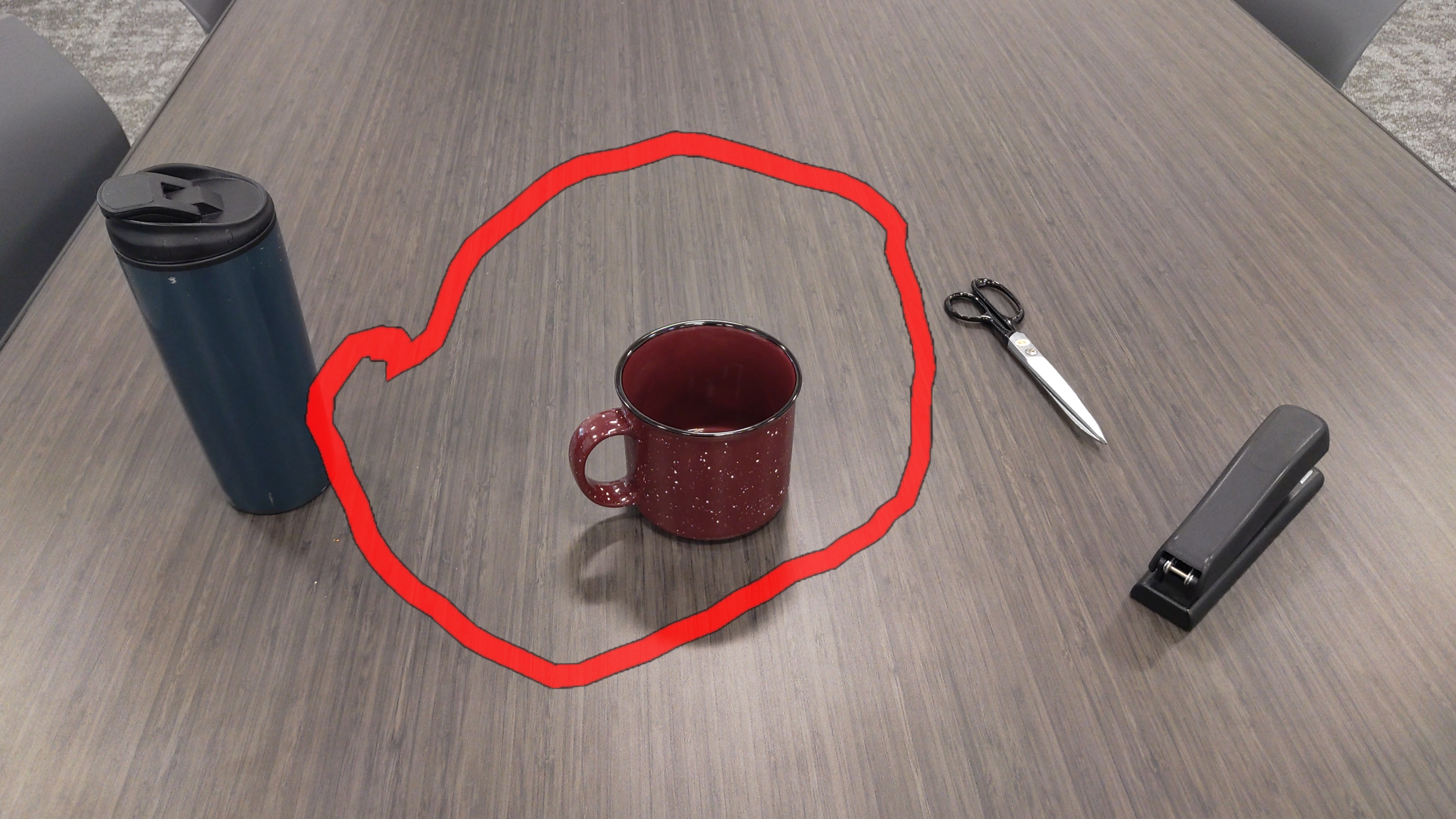}
    \label{fig:image3}
  \end{minipage}
  \begin{minipage}{0.24\textwidth}
    \includegraphics[width=\linewidth]{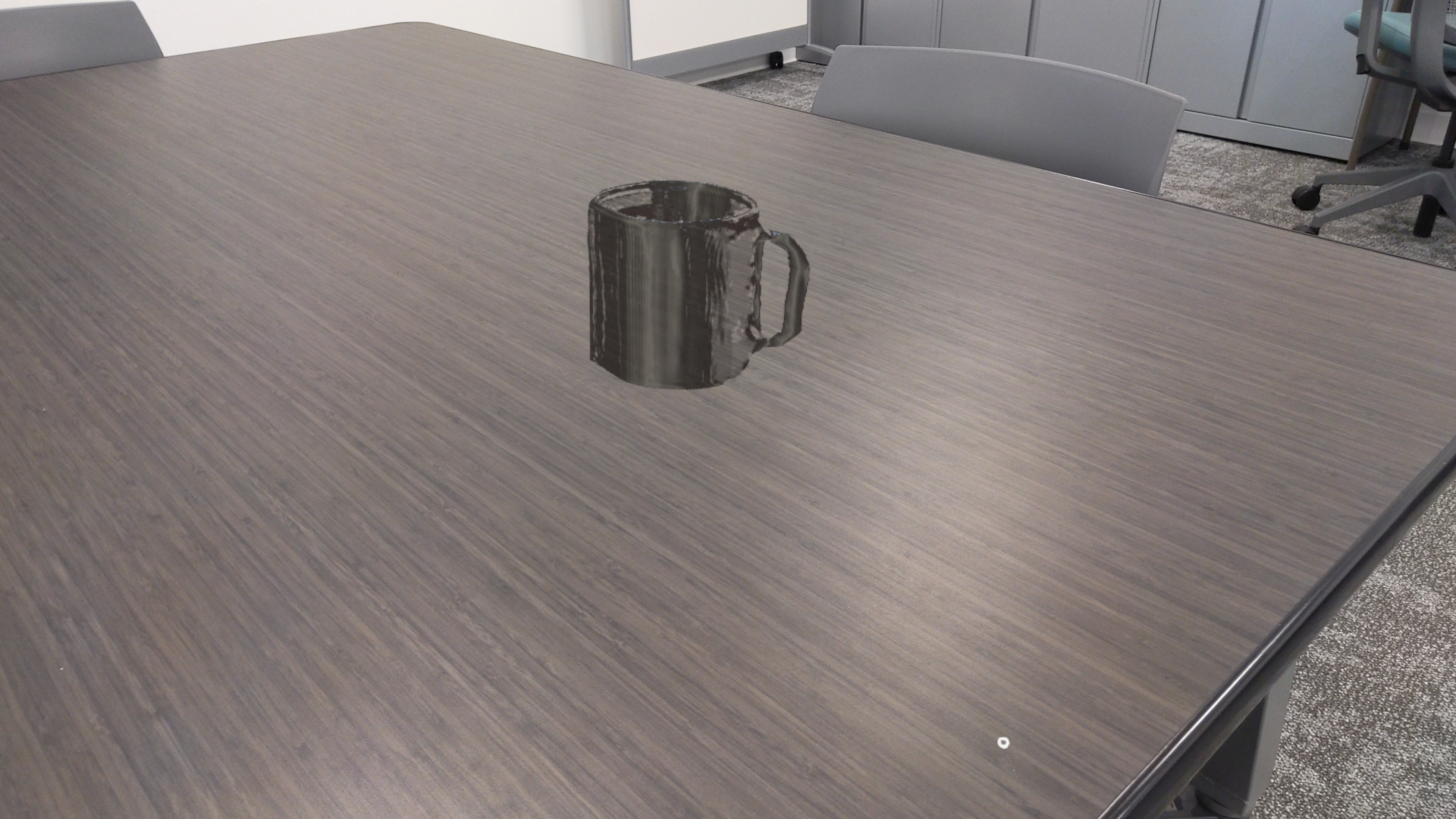}
    \label{fig:image4}
  \end{minipage}
  \caption{From a 2D image to an AR environment.
\textbf{Left half:} From a 2D image on a computer screen to a 3D object in an AR environment.
\textbf{Right half:} From a 2D image of a real world object to a 3D object in an AR environment.
}
\label{fig:ar_environment}
\end{figure*}

Ultimately, our goal is to push the boundaries of what is possible with generative AI and AR, creating a seamless and intuitive user experience.
By harnessing the power of these technologies, we aim to transform the way people interact with digital content, paving the way for a future where the physical and digital worlds coexist harmoniously (Figure~\ref{fig:ar_environment}).

\section{Related Works}
\label{sec:Related Works}

The landscape of image-to-3D generation has rapidly evolved, driven by advancements in generative AI technologies.
Researchers have developed a range of innovative methods to tackle the challenges of reconstructing 3D models from two-dimensional images, particularly focusing on single-view 3D reconstruction.
We provide an overview of significant contributions in this field.

\subsection{Image to 3D Generation}

The field of image-to-3D generation using generative AI has seen substantial progress, with numerous methodologies contributing to its rapid development.
We describe the key works and models that have advanced this domain, particularly in the context of single-view 3D reconstruction.

Early advancements in 3D generative modeling largely focused on the development and application of Generative Adversarial Networks (GANs)~\cite{goodfellow2014generative}.
This approach was pivotal in shaping the early landscape of 3D generation, enabling the creation of models from representations like point clouds~\cite{achlioptas2018learning} and implicit representations~\cite{chen2019learning}.
Building on this foundation, recent studies have integrated GANs with differentiable rendering methods, which utilize multiple rendered views as the basis for the loss signal, significantly enhancing the quality of generated 3D models~\cite{chan2022efficient}.

Building on these advancements, Behravan et al. proposed leveraging vision-language models (VLMs) for real-time 3D object generation in AR, emphasizing contextual understanding for enhanced object recommendation~\cite{BehravanVRST,behravan2024AIxVR}.
Similarly, Dasgupta et al. introduced a framework for real-time object recognition using edge computing, addressing hardware limitations in mixed reality~\cite{Dasgupta-2020-a}.
These approaches advance AI-driven decision-making and object interaction in AR and MR.

The advent of diffusion models marked a significant leap forward in high-quality image generation, with growing interest in extending these models to the 3D domain.
Typically, this involves training a Vector-Quantized Variational Autoencoder (VQ-VAE) on 3D representations such as triplane~\cite{shue2023triplane} and point cloud~\cite{zeng2022latent}, followed by applying the diffusion model in the latent space.
Although direct training on 3D representations is less common, it is gaining traction, particularly with point clouds~\cite{luo2021diffusion}, voxels \cite{zheng2023locally}, and neural wavelet coefficients~\cite{hui2022neural}.

Among the innovative methods, ShapeClipper stands out by reconstructing 3D object shapes from single-view RGB images using CLIP-based shape consistency and geometric constraints.
This approach leverages the semantic consistency of CLIP embeddings to enhance 3D shape learning, showing superior performance on datasets like Pix3D, Pascal3D+, and OpenImages~\cite{huang2023shapeclipper}.

Similarly, Shap-E employs autoencoders trained on explicit 3D representations combined with a diffusion model in the latent space, offering a flexible approach to representing complex 3D structures. 
This method addresses the limitations of fixed-resolution explicit outputs, providing a scalable solution for 3D model generation~\cite{achlioptas2018learning}.


Voxel-based methods have also seen significant advancements.
These approaches benefit from the regularity of voxel representation and the application of 2D convolutional neural networks (CNNs).
However, they often face a trade-off between resolution and computational efficiency.
For example, voxel-grid methods offer better accuracy than bounding boxes but at the cost of increased computational resources~\cite{zamir2018taskonomy}.
SparseFusion addresses this by integrating sparse voxel grids with neural networks, achieving high-fidelity 3D reconstructions from sparse input views, suitable for real-time AR and VR applications~\cite{fu2021auto}.


Implicit surface methods, including those utilizing neural radiance fields (NeRF), have gained prominence for their ability to implicitly represent complex geometries.
Techniques like PixelNeRF and CodeNeRF employ lightweight networks to model arbitrary topologies, proving effective for scene generalization and novel-view synthesis~\cite{yu2021pixelnerf}.




Lastly, the GET3D model by NVIDIA leverages recent advances in differentiable surface modeling and rendering, along with 2D Generative Adversarial Networks, to train a model that generates explicit textured 3D meshes.
This model excels in producing high-fidelity 3D shapes with detailed geometry and complex topology from 2D image collections, making it highly applicable across various real-world applications, from gaming and entertainment to e-commerce and design~\cite{gao2022get3d}.

These advancements highlight the transformative potential of generative AI in 3D model creation, democratizing access to high-quality 3D content and enabling innovative applications across AR.
\subsection{Image Processing and object detection}

Object detection and image processing have significantly evolved with the advent of deep learning and convolutional neural networks (CNNs).
Various approaches have been developed, each improving accuracy, speed, and robustness in different scenarios.

Early methods in object detection repurposed classifiers to perform detection.
Techniques like Deformable Parts Models (DPM) used a sliding window approach to run classifiers at evenly spaced locations over the entire image~\cite{ren2015faster}.
More recent approaches, such as R-CNN (Regions with CNN features), first generate potential bounding boxes in an image and then classify these proposed regions~\cite{girshick2014rich}.
This method laid the groundwork for many modern object detectors but was computationally expensive and slow.

YOLO (You Only Look Once) reframed object detection as a single regression problem, directly predicting bounding boxes and class probabilities from full images in one evaluation.
YOLO's unified architecture significantly sped up detection, making it possible to process images in real-time with high accuracy~\cite{redmon2016yolo,redmon2018yolov3}.

Faster R-CNN introduced Region Proposal Networks (RPNs) that share convolutional layers with detection networks, generating high-quality region proposals almost cost-free.
This method improved both the speed and accuracy of object detection, setting a new standard in the field~\cite{ren2015faster}.
The Faster R-CNN pipeline integrates region proposal and detection into a single, unified network, enhancing performance on various benchmarks~\cite{he2017mask}.

Another innovative approach is the CenterNet, which models an object as a single point — the center point of its bounding box.
This method uses keypoint estimation to find center points and regress to other object properties, achieving a balance of simplicity, speed, and accuracy~\cite{zhou2019objects}.

Sparse R-CNN and its variants, including methods like SpineNet and HitDetector, aim to further optimize the detection process by using sparse prediction techniques.
These methods focus on reducing the number of bounding boxes and improving the efficiency of the detection pipeline~\cite{peize2020sparse}.

Mask R-CNN extended Faster R-CNN by adding a branch for predicting segmentation masks on each region of interest, enabling pixel-level segmentation alongside object detection.
This advancement has been crucial for applications requiring detailed image understanding, such as scene parsing and instance segmentation~\cite{he2017mask,musyarofah2020mask,hassan2022review}.

The Least Absolute Shrinkage and Selection Operator (LASSO) method, widely used in statistical modeling and machine learning, has also found significant applications in image processing, particularly in the context of object detection and zone selection~\cite{tibshirani1996regression}.
Traditional LASSO techniques have been employed to enhance image segmentation by enforcing sparsity, which aids in isolating relevant features from complex backgrounds. This method ensures high-quality isolation and reconstruction of objects, addressing challenges related to complex scenes and backgrounds.
LassoNet, a deep neural network specifically designed for LASSO selection of 3D point clouds, exemplifies the effectiveness of combining LASSO techniques with deep learning to handle variability in data, viewpoints, and lasso selections, further enhancing the robustness and efficiency of the selection process~\cite{zhu2024lassonet}.

Overall, the development of these models and methods has paved the way for robust, real-time object detection systems that can handle complex backgrounds and multiple objects, addressing many challenges in image processing and object detection within AR environments.

\section{Challenges}

One of the primary challenges in converting images to 3D models within AR environments stems from the limitations of current AI models.
These models are typically designed to convert simple images, often with plain backgrounds, into 3D representations.
However, in AR, images are captured using a camera integrated into the headset, and these images often include complex backgrounds and multiple objects.
This complexity makes it difficult for the AI models to accurately identify and convert individual objects into 3D models, as the presence of multiple items and intricate backgrounds confuses the conversion algorithms.

\begin{figure}[!ht]
    \centering
    \begin{subfigure}[b]{0.45\textwidth}
        \centering
        \includegraphics[width=\textwidth]{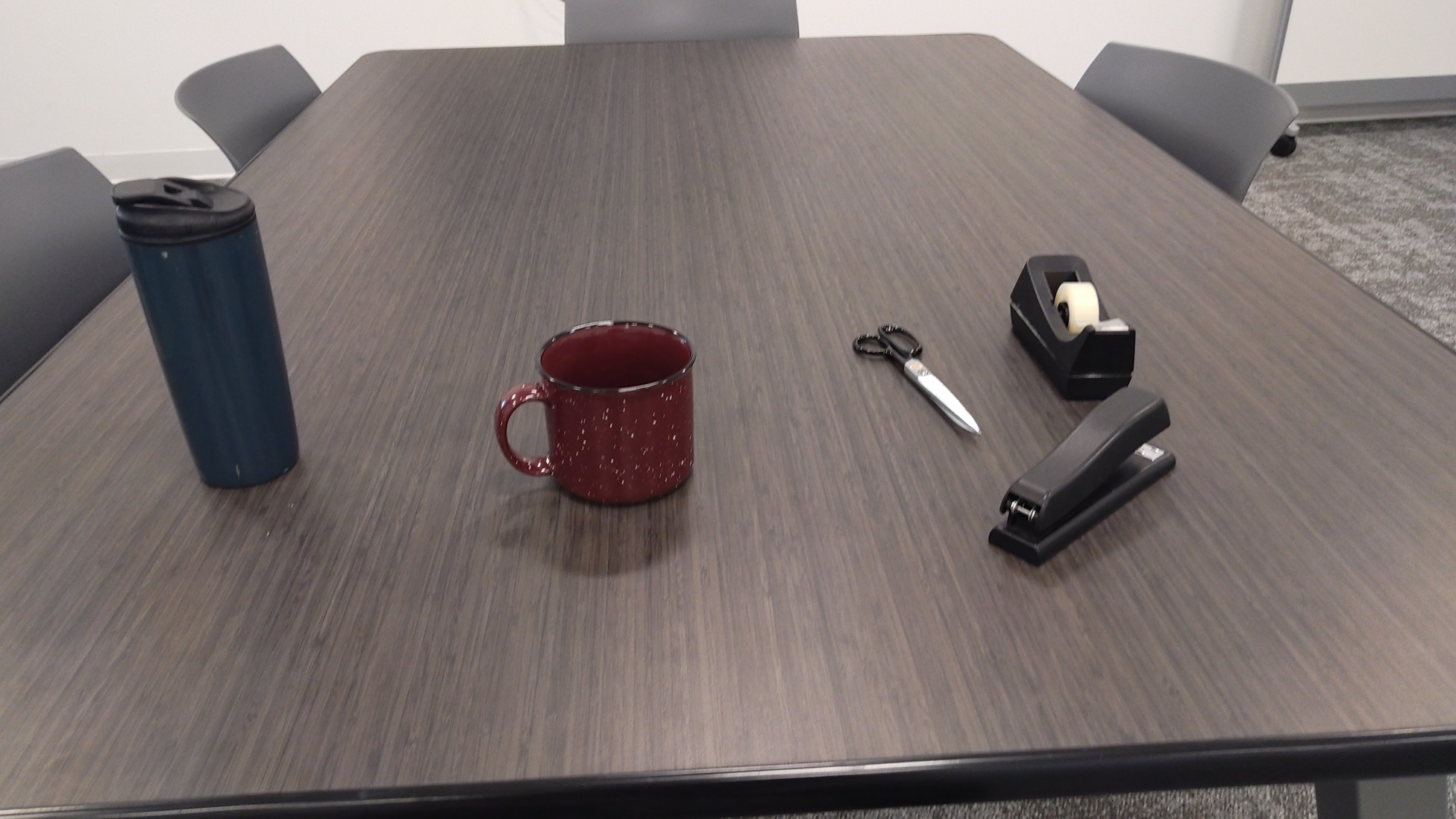}
        \caption{AR headset view showing multiple objects in a complex scene.}
        \label{fig:complex_scene}
    \end{subfigure}
    \hfill
    \begin{subfigure}[b]{0.45\textwidth}
        \centering
        \includegraphics[width=\textwidth]{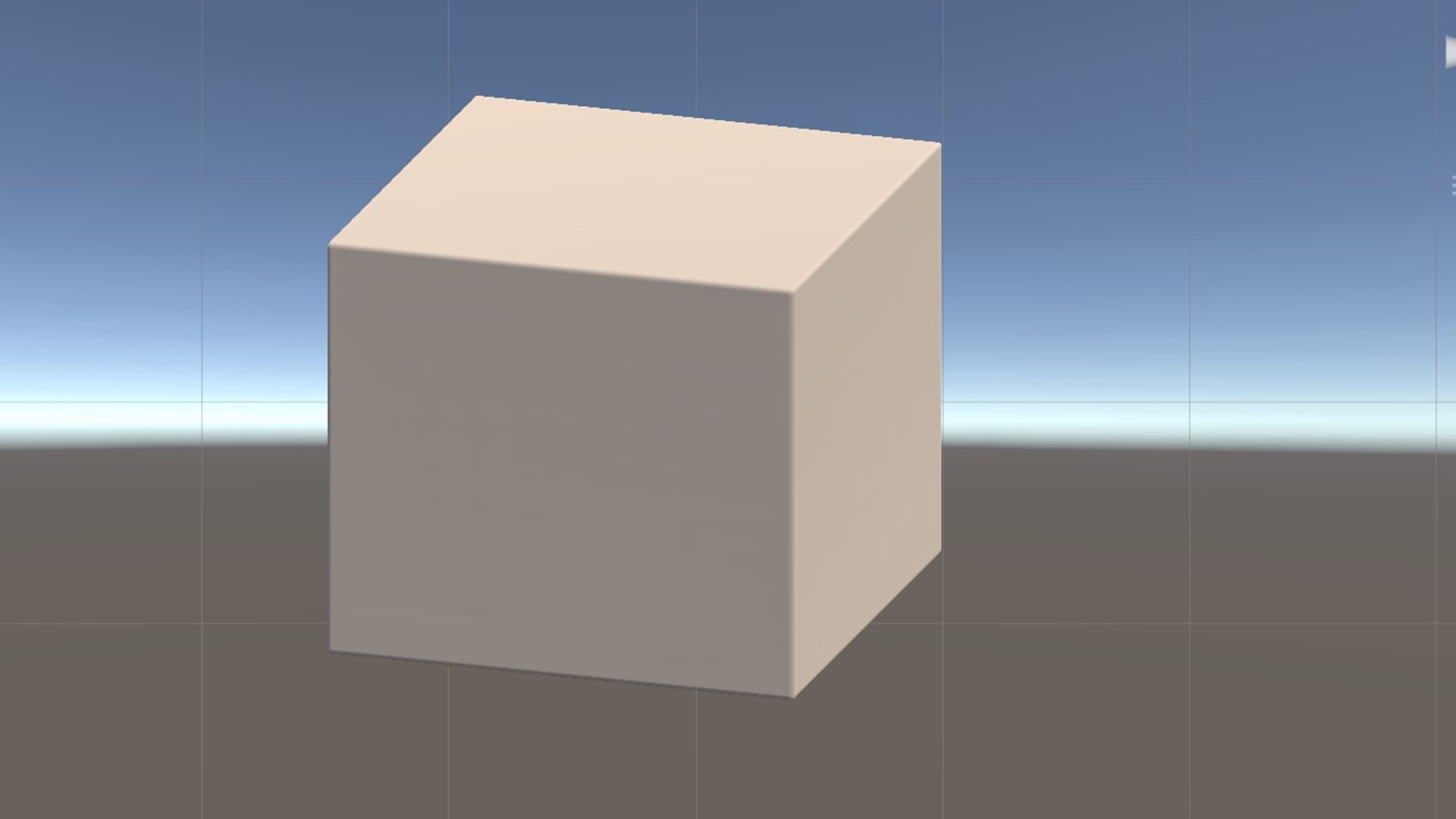}
        \caption{Shap-E model’s output for multiple objects in a complex scene.}
        \label{fig:multi_issue}
    \end{subfigure}
    
    \vspace{0.5cm}
    
    \begin{subfigure}[b]{0.45\textwidth}
        \centering
        \includegraphics[width=\textwidth]{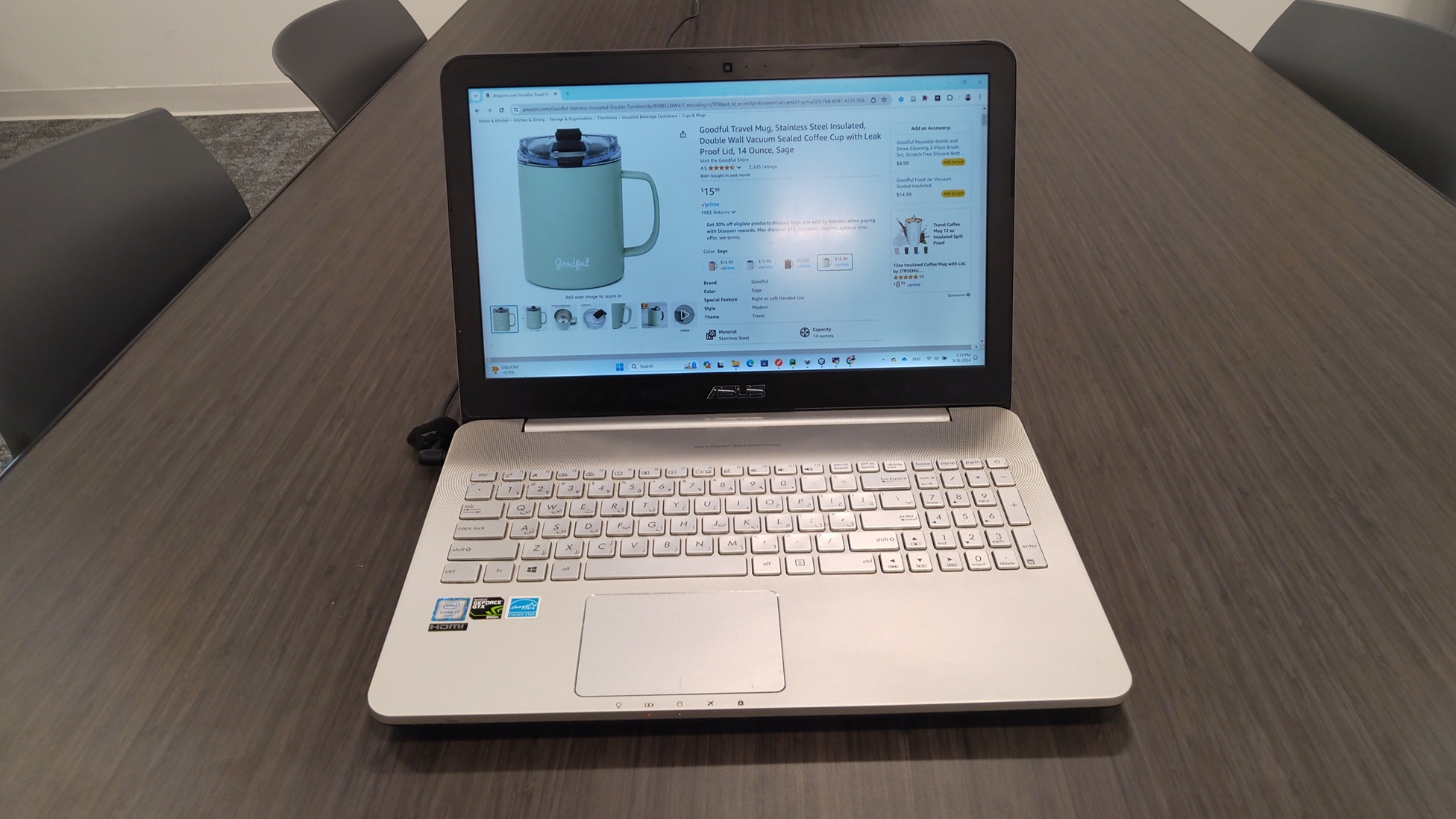}
        \caption{AR headset view showing a mug displayed on a monitor.}
        \label{fig:monitor_view}
    \end{subfigure}
    \hfill
    \begin{subfigure}[b]{0.45\textwidth}
        \centering
        \includegraphics[width=\textwidth]{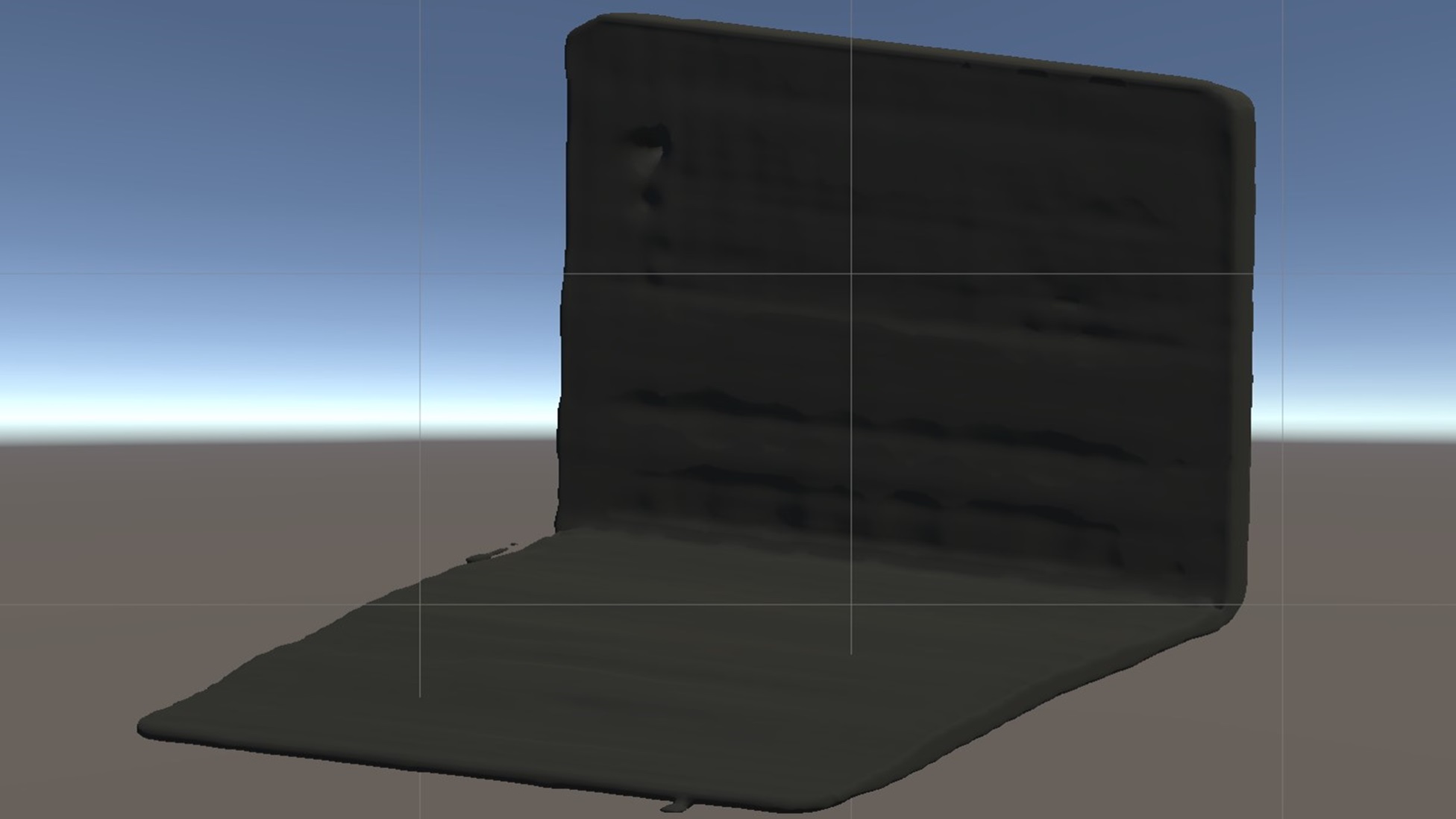}
        \caption{Shap-E model’s output for a mug displayed on a monitor.}
        \label{fig:monitor_issue}
    \end{subfigure}
    
    \caption{Challenges in AR headset views and Shap-E model outputs.}
\end{figure}

As shown in Figure~\ref{fig:complex_scene}, there are multiple objects present in the environment, and the image is captured from the view of the AR headset.
However, in  Figure~\ref{fig:multi_issue}, the output from the Shap-E model for this scene is just a single cube, indicating the model's inability to handle environments with multiple objects effectively.

Another significant challenge arises when trying to convert images of objects displayed on monitors, such as items from online stores, into 3D models.
When the user attempts to capture an image of a product on their screen, the resulting picture inevitably includes the monitor and surrounding elements.
Current image-to-3D conversion models are not equipped to isolate the object of interest from its background effectively.
Consequently, the output often includes the monitor itself as part of the 3D model, which is not the intended outcome.

As shown in Figure~\ref{fig:monitor_view}, a mug is displayed in an online store view captured by the AR headset.
Figure~\ref{fig:monitor_issue} shows the output from OpenAI's Shap-E model based on this image of the mug.
Interestingly, if the user intended to create the mug seen on the monitor, the model's output is actually a laptop.



In addition to these challenges, there is also the issue of varying lighting conditions and angles at which images are captured.
In real-world settings, lighting can significantly affect the appearance of objects, creating shadows, reflections, and other visual artifacts that can mislead the AI models.
These models often struggle to differentiate between actual object features and these artifacts, leading to inaccuracies in the generated 3D models.
Furthermore, the angle at which an image is taken can distort the perceived shape of the object, complicating the conversion process and resulting in 3D models that do not accurately reflect the original items.
 
Lastly, the integration of these 3D models into the AR environment itself poses a challenge.
Even if the conversion process is successful, the models must be seamlessly integrated into the user's real-time view.
This integration requires precise alignment and scaling to ensure that the virtual objects appear as part of the real world.
Any errors in this process can disrupt the user's experience and reduce the overall effectiveness of the AR application.
Therefore, advancements in both image-to-3D conversion and AR integration are crucial to overcoming these challenges and achieving seamless real-time content creation in AR environments.

 \section{System Workflow and AR Integration}

\begin{figure}
    \centering
    \includegraphics[width=1\linewidth]{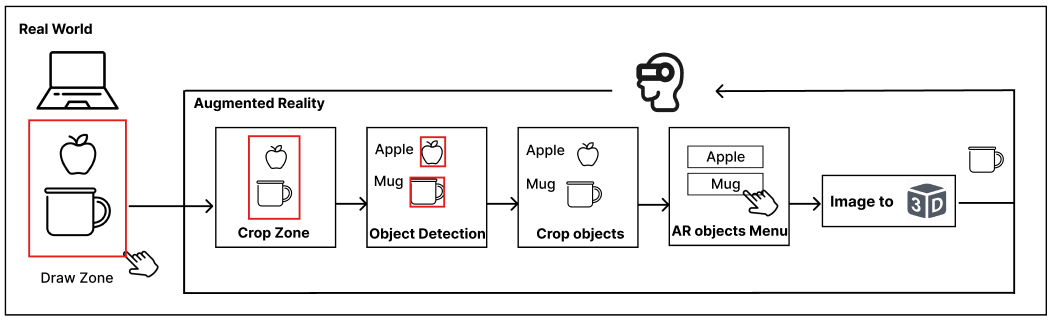}
    \caption{Cloning object workflow.}
    \label{fig:enter-label}
\end{figure}

Our primary contribution lies in the development of a process for converting 2D images into 3D models that can be seamlessly integrated into AR environments (Figure~\ref{fig:enter-label}).
Our approach focuses on enhancing the capabilities of generative AI models to achieve this conversion.
We have meticulously designed a series of steps, each tailored to address the specific challenges and limitations identified in our previous research.
By doing so, we ensure that the generated 3D models are not only accurate and detailed but also optimized for real-time application in AR. 


\subsection{Zone Selection and Image Capture}

\begin{figure}
    \centering
    \includegraphics[width=0.6\linewidth]{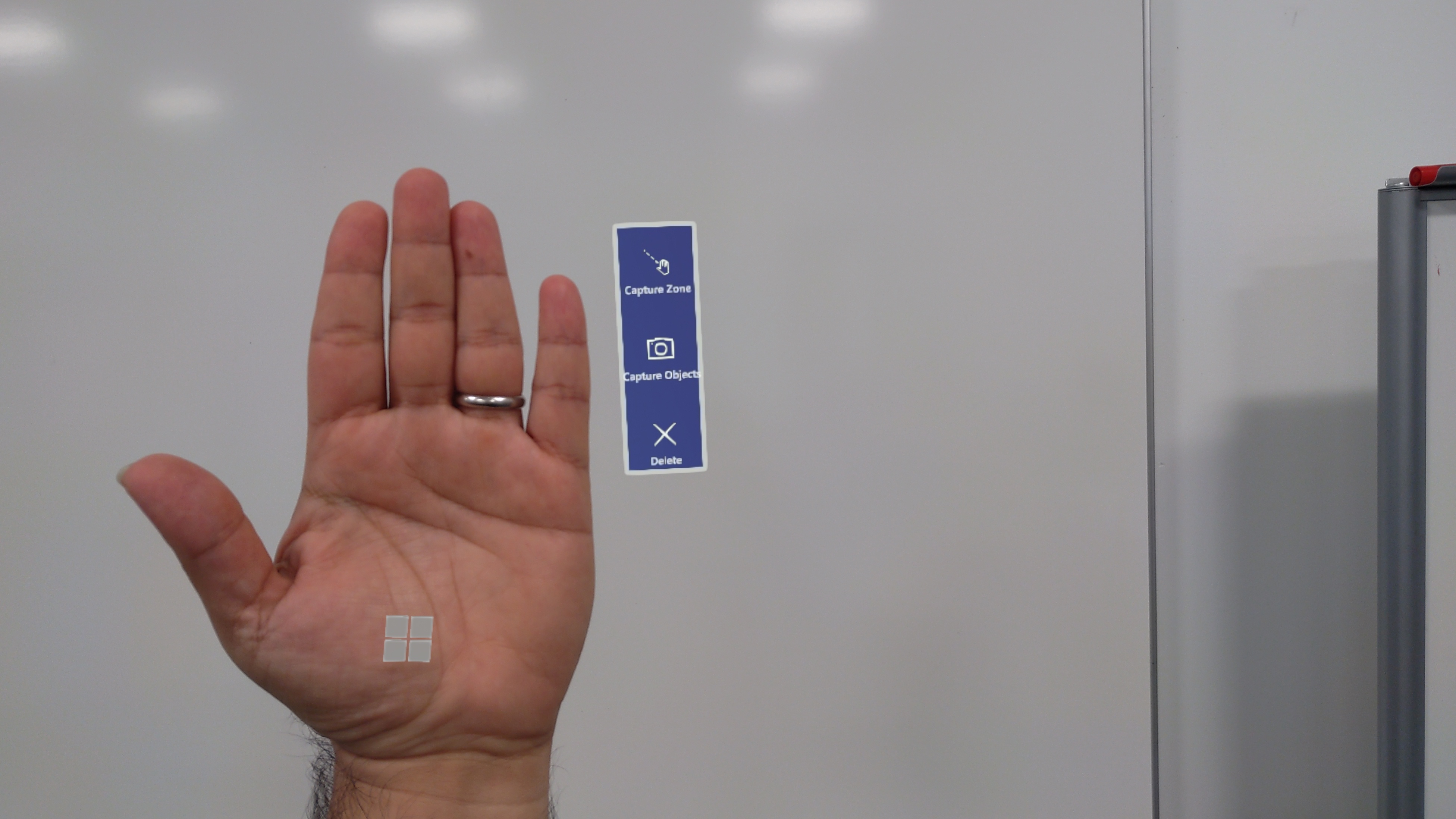} 
    \caption{AR hand menu.}  
    \label{fig:Hand_Menu}
\end{figure}

\begin{figure}[!ht]
\centering
\includegraphics[width=0.6\linewidth]{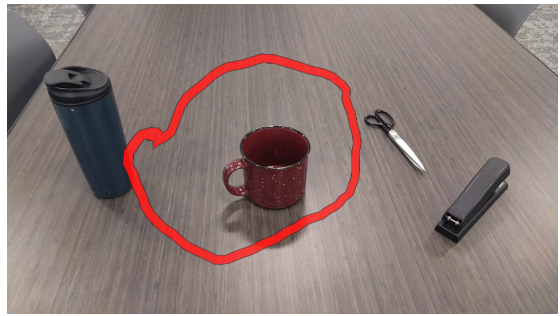}  
\caption{Drawing selection zone (Lasso selection interaction technique).}  
\label{fig:Drawing}
\end{figure}

In the AR hand menu shown in Figure~\ref{fig:Hand_Menu}, users can click on the ``capture zone'' button to draw a red line around the object or objects they want to generate 3D models of in the AR environment.
This lasso selection interaction technique is used in 2D and 3D settings~\cite{Yu-2016-a}.
The interactive process, illustrated in Figure~\ref{fig:Drawing}, allows for precise zone selection, ensuring that the desired objects are accurately isolated from the background.
The process of drawing the line is stopped by pinching, and after three seconds, an image is captured, including the line drawn by the user and the objects within the user's view.

Additionally, users can click on the ``capture objects'' button in the hand menu to capture an image and request the creation of 3D models for all objects in their view without drawing a line.

\subsection{Crop Selected Zone}

\begin{figure}[!ht]
\centering
\includegraphics[width=0.4\linewidth]{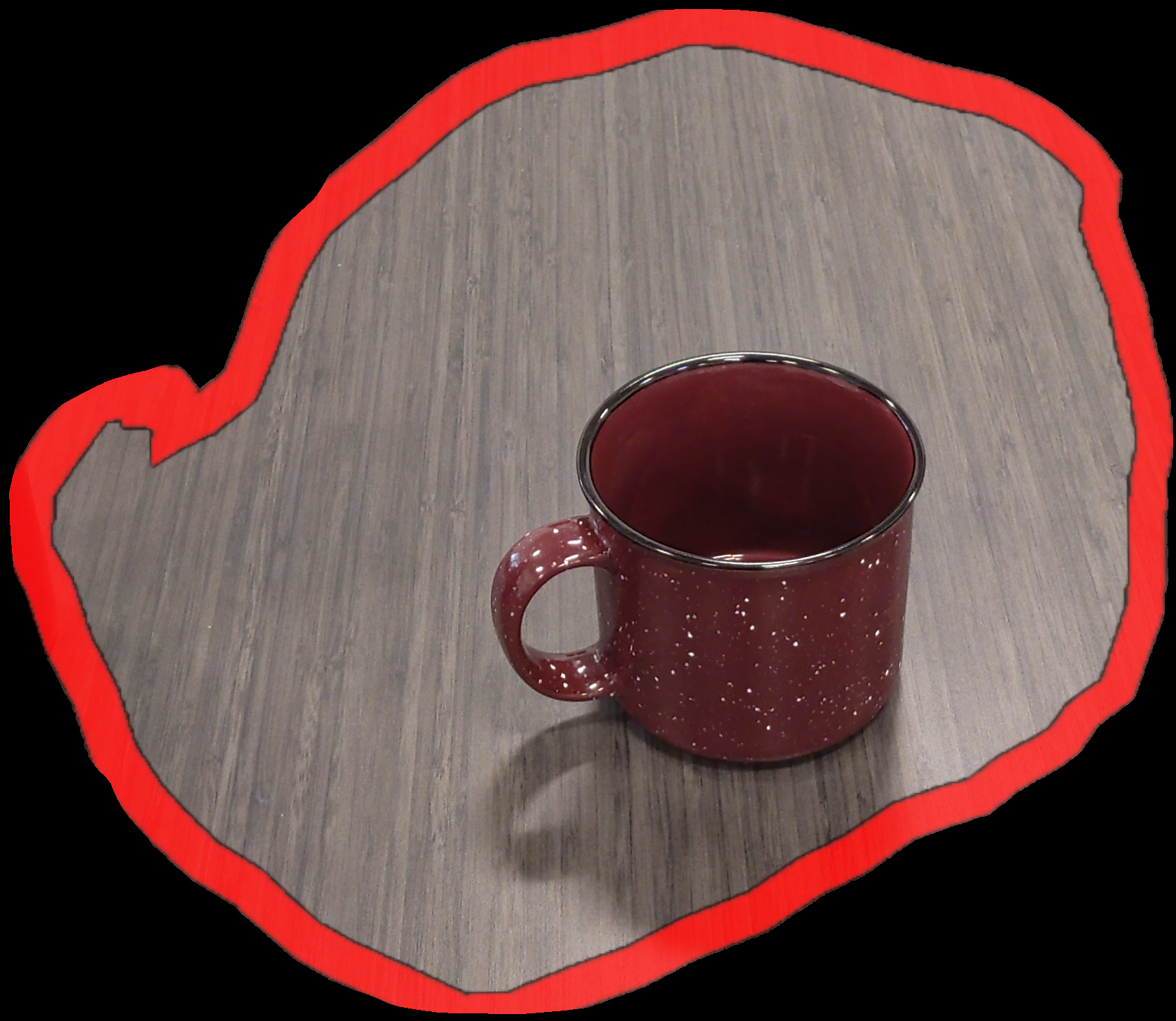} 
\caption{Crop selected zone.}  
\label{fig:Crop}
\end{figure}

The selected zone is then cropped to focus on the area of interest, as seen in \ref{fig:Crop}.
This cropping helps in reducing the computational load and improves the model's focus on the relevant part of the image.
This is achieved by combining the user's actions within the AR environment with the cropping process.
 
\subsection{Object Detection}
 
\begin{figure}[!ht]
\centering
\includegraphics[width=0.4\linewidth]{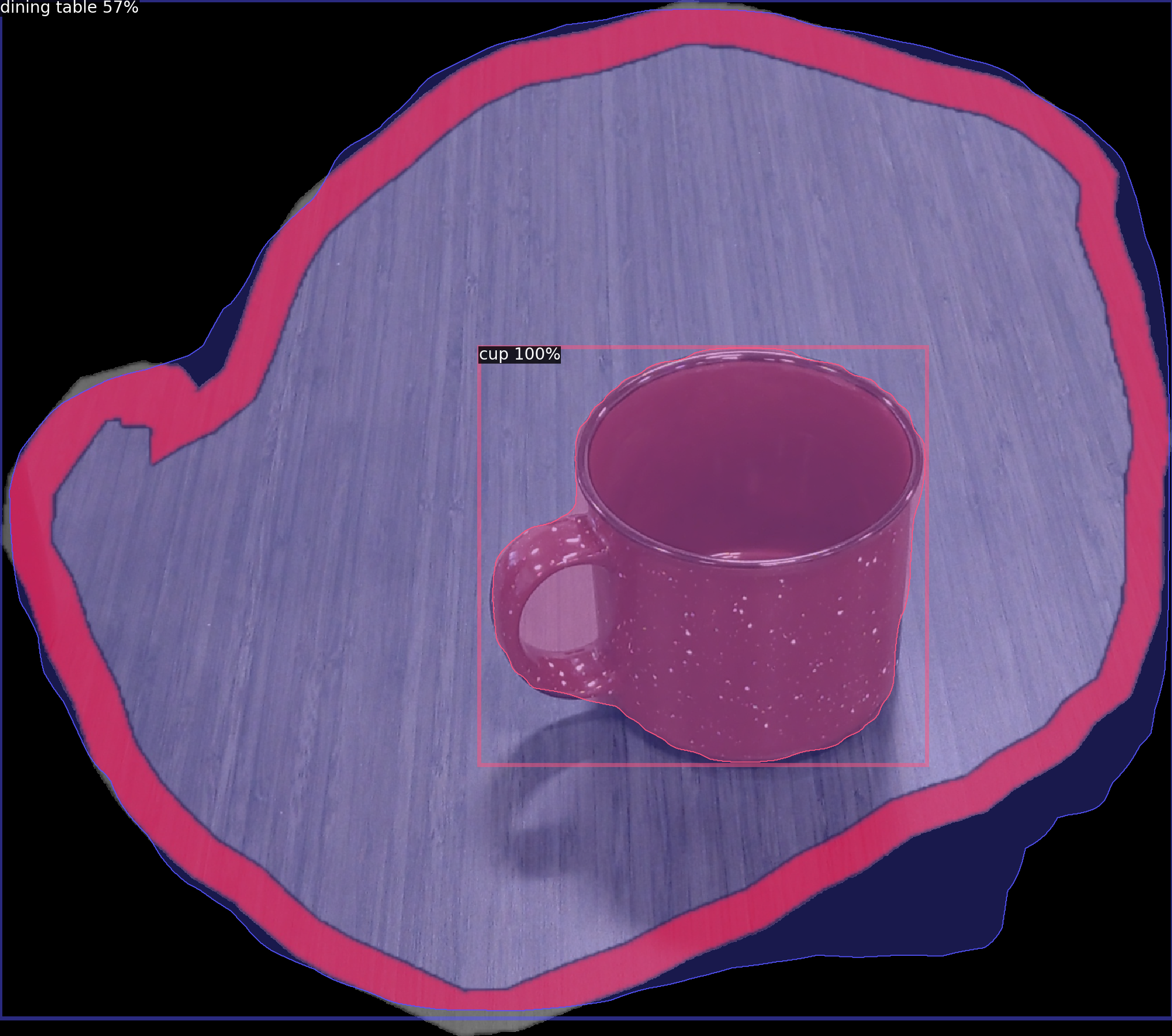} 
\caption{Object detection in zone.} 
\label{fig:Obj_detection}
\end{figure}

Advanced object detection algorithms are employed to identify and outline objects within the selected zone, as illustrated in Figure~\ref{fig:Obj_detection}.
This step is crucial in ensuring the accuracy and robustness of the object detection process, which subsequently influences the quality of the generated 3D models.
For this purpose, we utilized the Mask R-CNN (Region-based Convolutional Neural Networks) algorithm, specifically configured for instance segmentation.
The Mask R-CNN model is well-regarded for its ability to perform pixel-level segmentation, enabling precise detection and delineation of objects within an image.

To optimize the performance of our detection model, we utilized pre-defined settings tailored for instance segmentation tasks.
These settings include parameters that control various aspects of the model's operation.
One critical parameter is the confidence threshold for making predictions, which we set at 0.5.
This threshold determines the minimum confidence score required for an object to be considered a valid detection.
By fine-tuning this parameter, we ensured that our model strikes a balance between precision and recall, reducing the likelihood of false positives while maintaining high detection accuracy.

Another critical aspect of our approach involves the use of pre-trained weights for the detection model.
These weights, derived from extensive training on large datasets such as the COCO dataset, provide a robust foundation that enhances the model's ability to recognize and segment objects accurately.
Leveraging these pre-trained weights allows our model to benefit from the knowledge encoded during training, thereby improving its performance on new, unseen images.
This approach ensures that our detection process is both efficient and effective, capable of handling the variability and complexity of real-world scenes.

The integration of the detection model with the AR system is designed to utilize available computational resources optimally.
By automatically selecting the most suitable hardware, such as GPUs when available, the system ensures efficient processing and faster detection times.
This optimization is crucial for real-time applications, where timely and accurate detection is essential for a seamless user experience.
Through this integration, our system provides robust and reliable object detection capabilities, forming the foundation for subsequent 3D model generation and AR interactions.

\begin{figure}[!ht]
\centering
\includegraphics[width=0.4\linewidth]{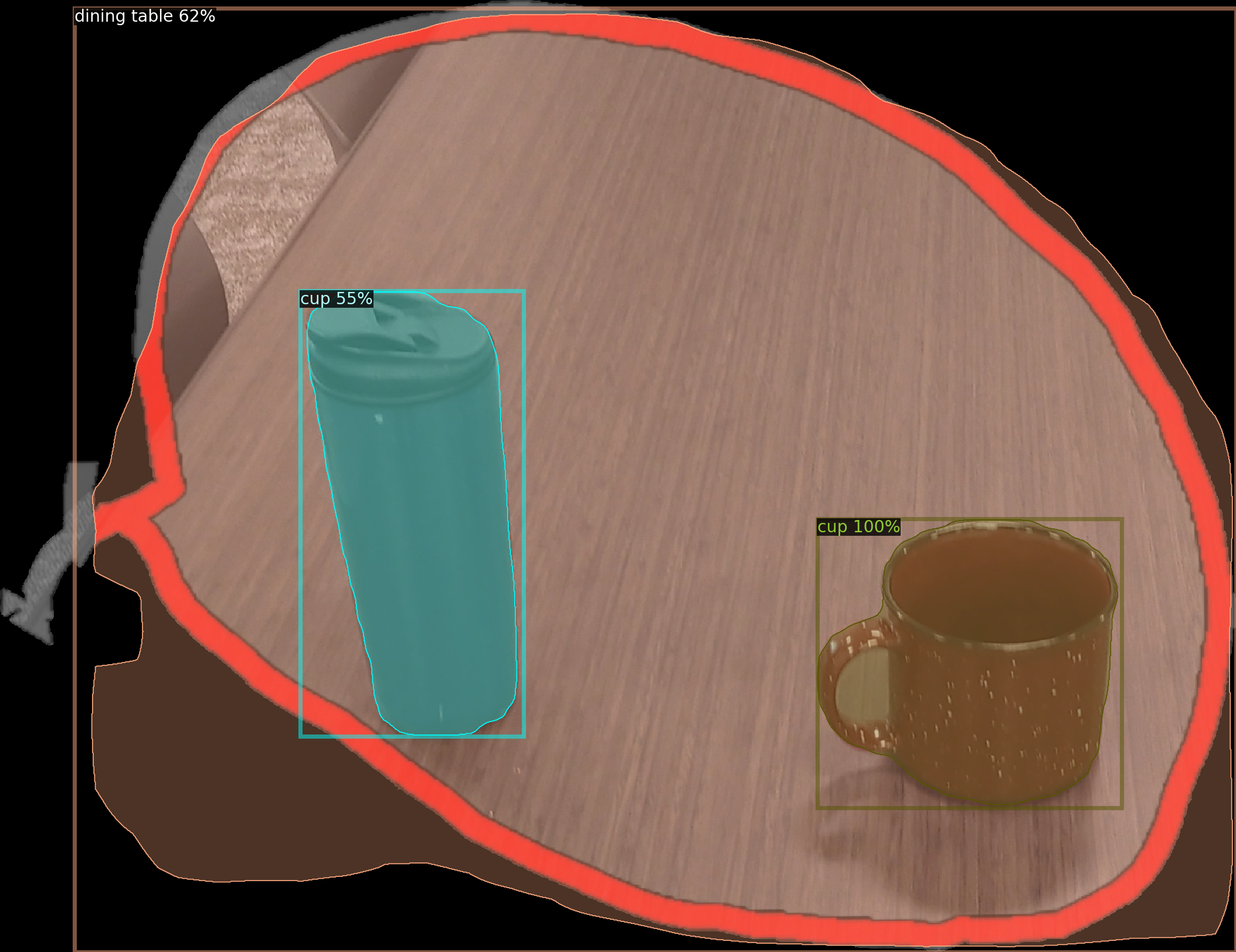} 
\caption{Multiple object detection in zone.}  
\label{fig:Multiobj_inZone}
\end{figure}

For scenes with multiple objects, the detection algorithm differentiates between various items, ensuring each object is accurately identified and processed separately, as depicted in Figure~\ref{fig:Multiobj_inZone}.
This capability is essential for creating detailed and accurate 3D models, as it ensures that each object within the scene is individually recognized and reconstructed.
By employing advanced object detection techniques with well-configured parameters, our approach ensures high precision and reliability, facilitating the seamless integration of virtual objects into the real world and enhancing the overall AR experience.

\begin{figure}[!ht]
\centering
\includegraphics[width=0.6\linewidth]{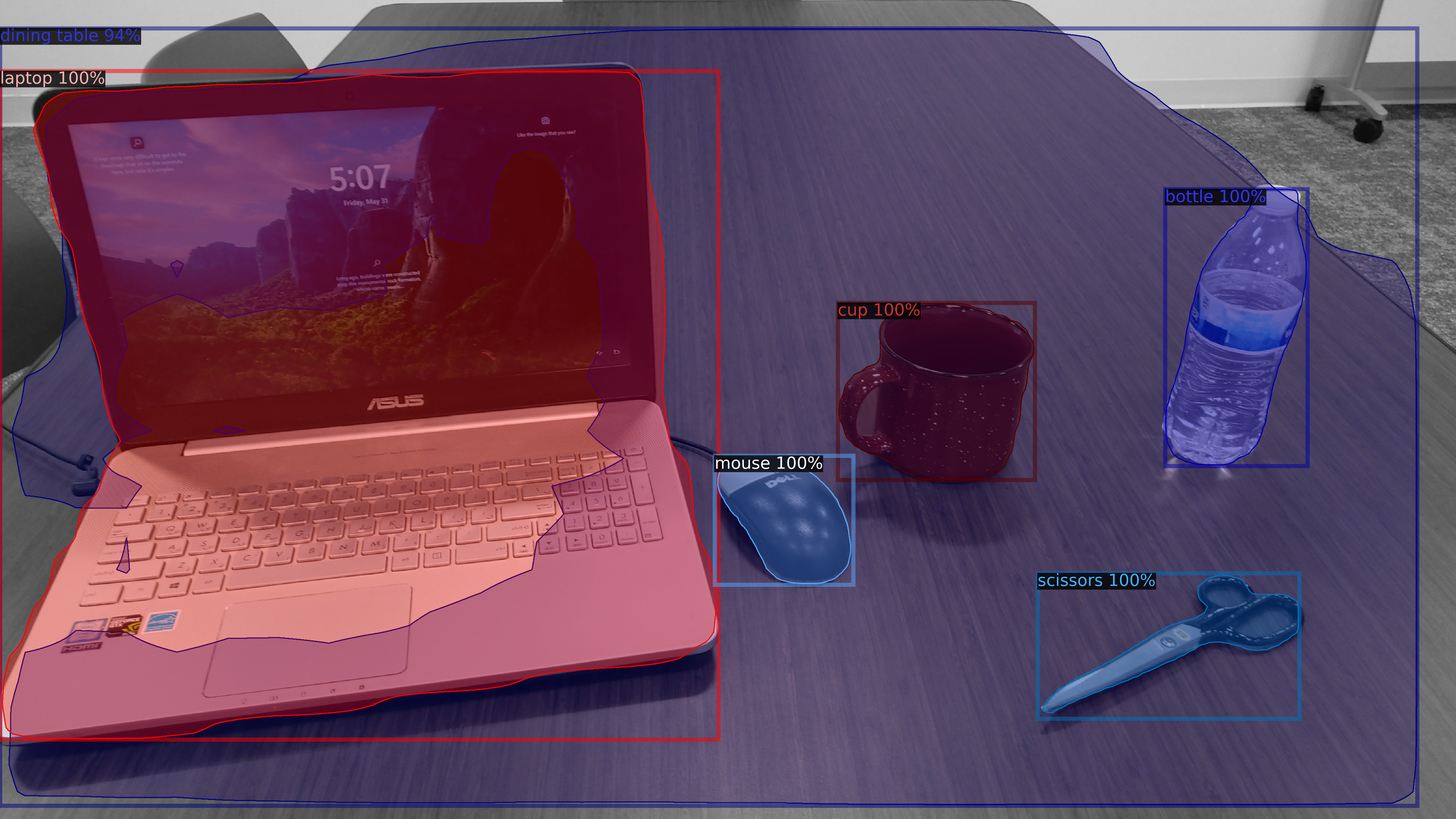} 
\caption{Capture all objects.}  
\label{fig:Multiobj_Capture All}
\end{figure}

If the user has utilized the ``capture objects'' button, similar to Figure~\ref{fig:Multiobj_Capture All}, all objects will be detected.

\subsection{Objects Cropping}

After detecting the objects within the selected zone, each object is precisely cropped around the edges to isolate it from the background.
To achieve this, we utilize a combination of contour detection and Mask R-CNN for instance segmentation.
Initially, the image is converted to the HSV color space to detect specific color ranges, such as red, to create a mask.
Contours within the mask are identified, and the largest contour is presumed to be the boundary of the desired zone.
This contour is used to create a mask isolating the object from the background.

Using OpenCV, the bounding box around the largest contour is calculated to crop the object.
This cropped region undergoes further processing using Mask R-CNN to detect and segment individual objects within the cropped zone.
The detected objects' masks and bounding boxes are used to extract precise regions corresponding to each object.
These regions are then labeled with their respective class names, and the objects are encoded to base64 format for efficient data transfer via API.
This technical approach ensures high accuracy in object cropping, facilitating seamless integration with generative AI models for 3D model generation.

\begin{figure}[!ht]
\centering
\includegraphics[width=0.4\linewidth]{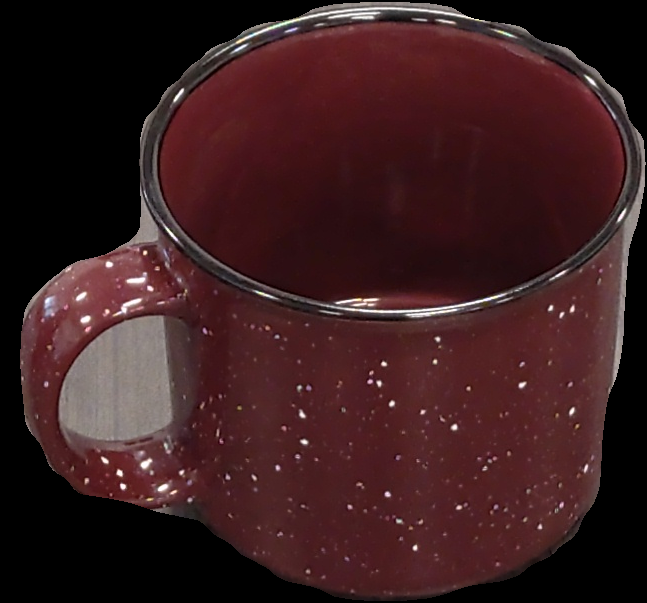} 
\caption{Object precisely cropped around the edges.}  
\label{fig:Croped_obj}
\end{figure}

As shown in Figure~\ref{fig:Croped_obj}, this step involves detailed edge cropping to obtain a clean object image.
Once cropped, these objects are labeled with their respective names and prepared for integration.
The final output, consisting of the cropped objects along with their names, is then sent via an API.
This integration allows for seamless communication between the image processing system and the generative AI models, facilitating efficient and accurate 3D model generation.

\subsection{User Verification and Object Selection}

\begin{figure}[!ht]
\centering
\includegraphics[width=0.6\linewidth]{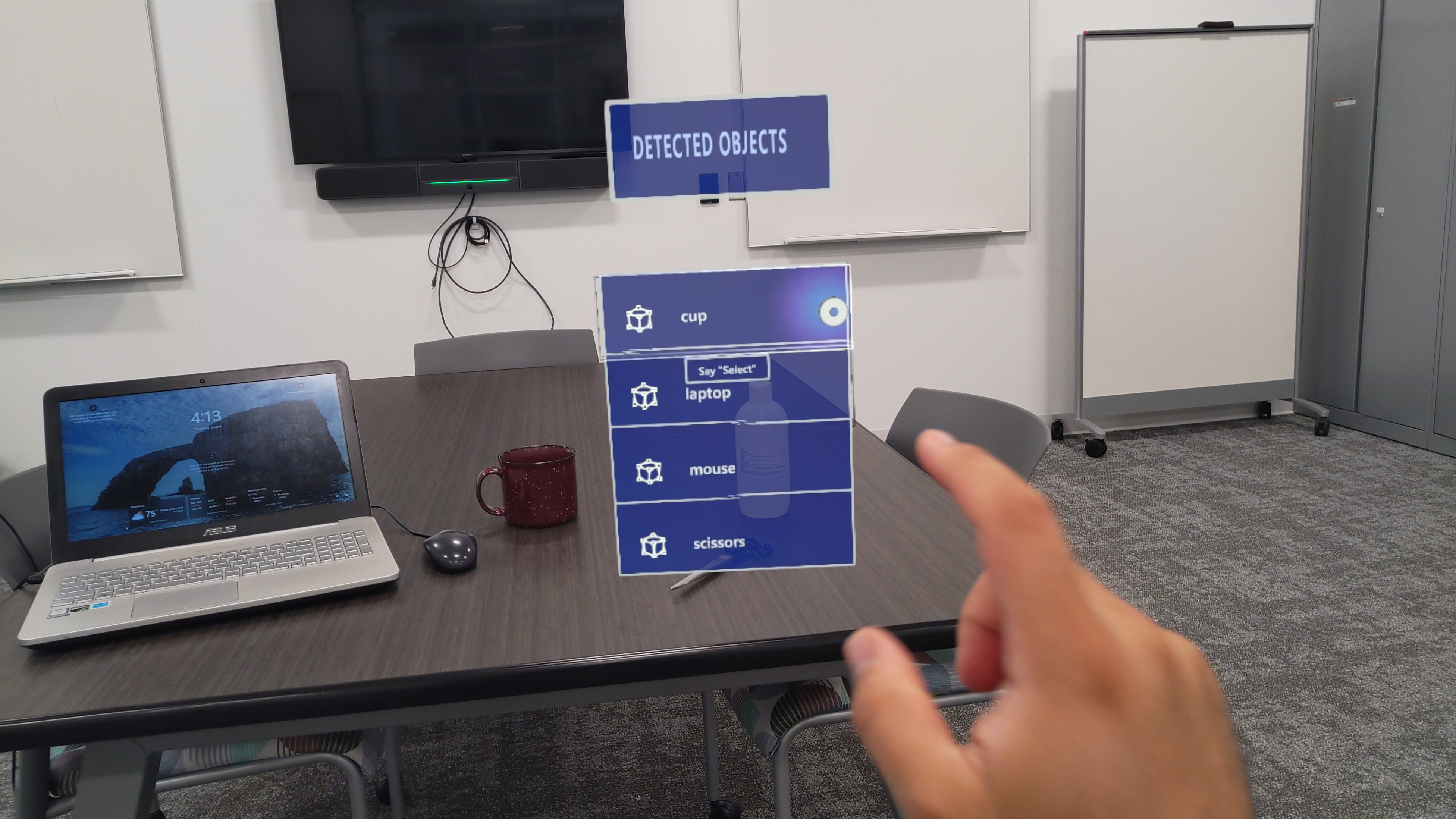} 
\caption{Detected objects menu.}  
\label{fig:objects_menu}
\end{figure}

The list of extracted objects from the previous step is presented to the user in the form of a menu (Figure~\ref{fig:objects_menu}).
This interface allows the user to review the detected objects and prevents the creation of any inaccurately identified items.
The user can select objects based on their priorities and preferences for conversion into 3D models.
This verification process enhances the overall reliability and user control within the AR environment.

\subsection{3D Model Generation and Display}

\begin{figure}[!ht]
\centering
\includegraphics[width=0.4\linewidth]{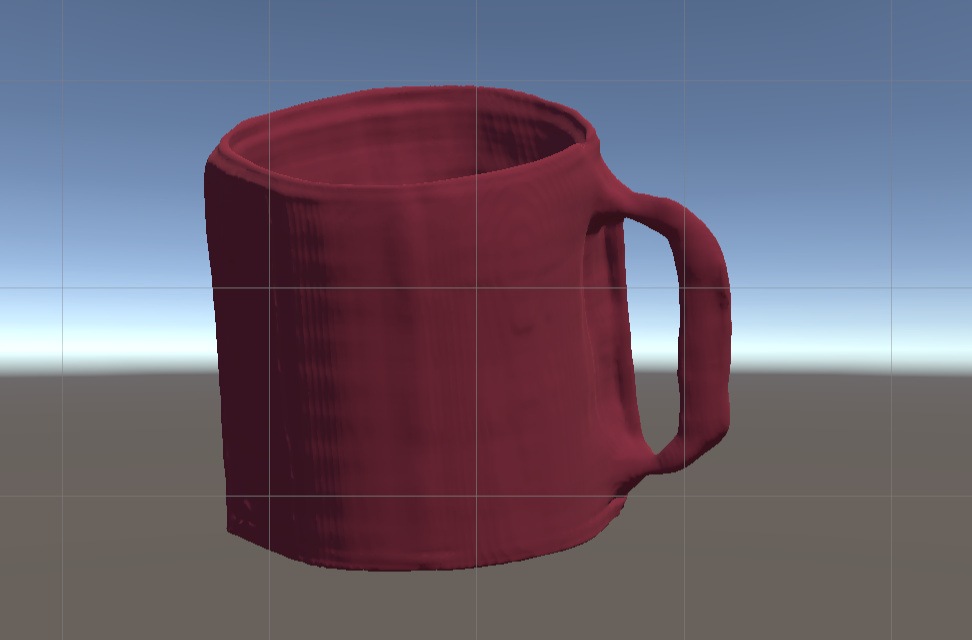} 
\caption{3D model generated by Shap-E.}  
\label{fig:output}
\end{figure}

Based on the user's selection from the menu, the extracted image is sent to the Shap-E model for 3D model generation.
The Shap-E model processes the image and creates a corresponding 3D model, which is then displayed to the user.
This step ensures that the user can visualize the generated 3D model and make any necessary adjustments.
The output from the Shap-E model is shown in Figure~\ref{fig:output}.

\section{Performance Evaluation}
To assess the performance of the Image-to-3D subsystem within the Matrix framework for real-time 3D object generation in AR environments, we evaluated two main phases: \textit{image processing for object detection} and \textit{image-to-3D conversion}. These phases were analyzed for their time efficiency and GPU usage, which are critical metrics in determining the system’s suitability for real-time AR applications.

\subsection{Key Metrics}

\begin{itemize}
    \item \textbf{Image Processing for Object Detection Time}: This metric measures the time required to detect objects within a captured image, identify and list their names, and crop the objects in the image as necessary before proceeding with 3D conversion. Efficient object detection is crucial to minimizing delays in real-time AR interactions.

    \item \textbf{Image-to-3D Conversion Time}: This metric measures the time taken to convert a detected object from a 2D image into a 3D model that can be rendered in an AR environment.

 \item \textbf{Vertices Reduction and Rendering Time}: This metric captures the time required to simplify the 3D model by reducing the number of vertices for efficient storage, download, and rendering. It includes:
    \begin{enumerate}
        \item \textbf{Model Simplification Time}: The time to process and reduce the 3D model's complexity while maintaining visual fidelity.
        \item \textbf{Load and Render Time}: The duration to download the reduced 3D model and render it in the AR environment. Optimizing this metric is critical for reducing latency and ensuring smooth user interactions, particularly in resource-constrained or mobile AR scenarios.
    \end{enumerate}

    \item \textbf{GPU Utilization}: GPU resource usage during the object detection and 3D conversion phases. Reduced GPU load is crucial in AR applications with limited hardware resources.
    
\end{itemize}

\subsection{Performance Summary}

Table \ref{tab:performance_metrics} summarizes the performance metrics for the Image-to-3D subsystem, with separate metrics for image processing (object detection) and image-to-3D conversion.

\begin{table}[h!]
\centering
\caption{Performance metrics for Image-to-3D subsystem.}
\label{tab:performance_metrics}
\begin{tabular}{@{}lcc@{}}
\toprule
\textbf{Metric} & \textbf{Measured Value} \\ \midrule
Image Processing for Object Detection Time (s) & 5.2 \\
Image-to-3D Conversion Time (s)                & 43.2  \\
Model Simplification Time (s)                  & 9.1   \\
Load and Render Time (s)                       & 10.3  \\
Average GPU Utilization (\%)                   & 61\% \\
GPU Memory Consumption (GB)                    & 6.8 \\ \bottomrule
\end{tabular}
\end{table}

The evaluation of the Image-to-3D subsystem highlighted the benefits of optimized object detection and mesh size reduction during image-to-3D conversion. These improvements contribute to a more resource-efficient framework, ensuring smoother user interactions in AR environments with limited computational resources.

\subsection{User Study}

The study included 35 participants from diverse backgrounds with varying levels of experience in AR/VR technologies, ensuring a broad range of perspectives. Participants ranged in age from 18 to 40, with a gender distribution of predominantly male (63\%) and female (37\%), alongside additional inclusive categories. AR/VR familiarity varied, with a majority reporting rare or no usage, and a small proportion being regular users. Familiarity with 3D design tools was predominantly slight or non-existent, with fewer participants having moderate to advanced expertise. Comfort with AR headsets spanned from “Never used” to “Very comfortable,” with the largest group moderately comfortable. This diversity facilitated a comprehensive evaluation of the system’s usability across demographic and experiential dimensions.

\subsection{Evaluation of System Usability Scale}

The SUS was used to evaluate user perceptions of the system's usability. The SUS provides an overall score based on users' responses to ten statements, alternating between positively and negatively worded items. Responses were collected from participants, with scores for each question adjusted to align with the SUS scoring guidelines. Specifically, scores for positively worded statements were adjusted by subtracting 1 from score, while scores for negatively worded statements were adjusted by subtracting score from 5. Each participant’s total score was then multiplied by 2.5 to convert it to a scale of 0-100.

The average SUS score across all participants was \textbf{69.64}, which is slightly above the established average threshold of 68 for usability. This score indicates that the system is perceived as usable but suggests areas for further enhancement. Specifically, the score reflects:

\begin{itemize}
    \item \textbf{Frequent Use and Ease of Learning}: Responses to statements like ``I would like to use this system frequently'' and ``I would imagine that most people would learn to use this system very quickly'' suggest that users generally view the system favorably regarding routine use and ease of learning. This is reflected in relatively higher scores for these items, supporting the system’s accessibility and appeal for regular use.

    \item \textbf{Complexity and Need for Support}: Lower scores on items related to complexity, such as ``I found the system unnecessarily complex,'' and the perception of needing technical support indicate that some users may find certain aspects of the system challenging to navigate. This suggests that while the system is overall usable, there may be specific functions or interfaces that could benefit from simplification or additional support features.

    \item \textbf{Confidence in Use}: Users generally felt confident using the system, as indicated by their responses to ``I felt very confident using the system.'' This positive feedback highlights the system's success in creating an environment where users feel competent and secure in their interactions.
\end{itemize}

To further analyze the results, participants were divided into two groups based on their frequency of use: \textbf{Rarely/Never} and \textbf{Regularly/Sometimes/Often}. An ANOVA test was conducted to determine whether there was a statistically significant difference in SUS scores between these two groups. 

\subsubsection{ANOVA Analysis}

The results of the ANOVA test indicated a significant difference in SUS scores between the two groups (\textit{F} = 18.212, \textit{p} < 0.001). The mean SUS score for the \textbf{Rarely/Never} group was \textbf{64.38}, while the mean SUS score for the \textbf{Regularly/Sometimes/Often} group was significantly higher at \textbf{80.71}, as shown in Table \ref{table:anova_sus}.

\begin{table}[h!]
\centering
\caption{Detailed ANOVA calculations for SUS (System Usability Scale).}
\label{table:anova_sus}
\begin{tabular}{|l|c|}
\hline
\textbf{Statistic} & \textbf{Value} \\ \hline
Group 1 (Rarely/Never) Mean Score & 64.38 \\ \hline
Group 2 (Regularly/Sometimes/Often) Mean Score & 80.71 \\ \hline
Group 1 (Rarely/Never) Variance & 50.32 \\ \hline
Group 2 (Regularly/Sometimes/Often) Variance & 42.17 \\ \hline
Number of Participants in Group 1 & 20 \\ \hline
Number of Participants in Group 2 & 15 \\ \hline
F-Statistic & 18.21 \\ \hline
P-Value & 0.000032 \\ \hline
\end{tabular}
\end{table}

\begin{itemize}
    \item \textbf{Rarely/Never Group}: Participants in this group rated the system lower on average, suggesting usability challenges that may discourage frequent use.
    \item \textbf{Regularly/Sometimes/Often Group}: This group had higher average scores, indicating a more favorable perception of the system's usability among participants who engage with it more frequently.
\end{itemize}

\begin{figure}[!ht]
    \centering
    \includegraphics[width=0.7\textwidth]{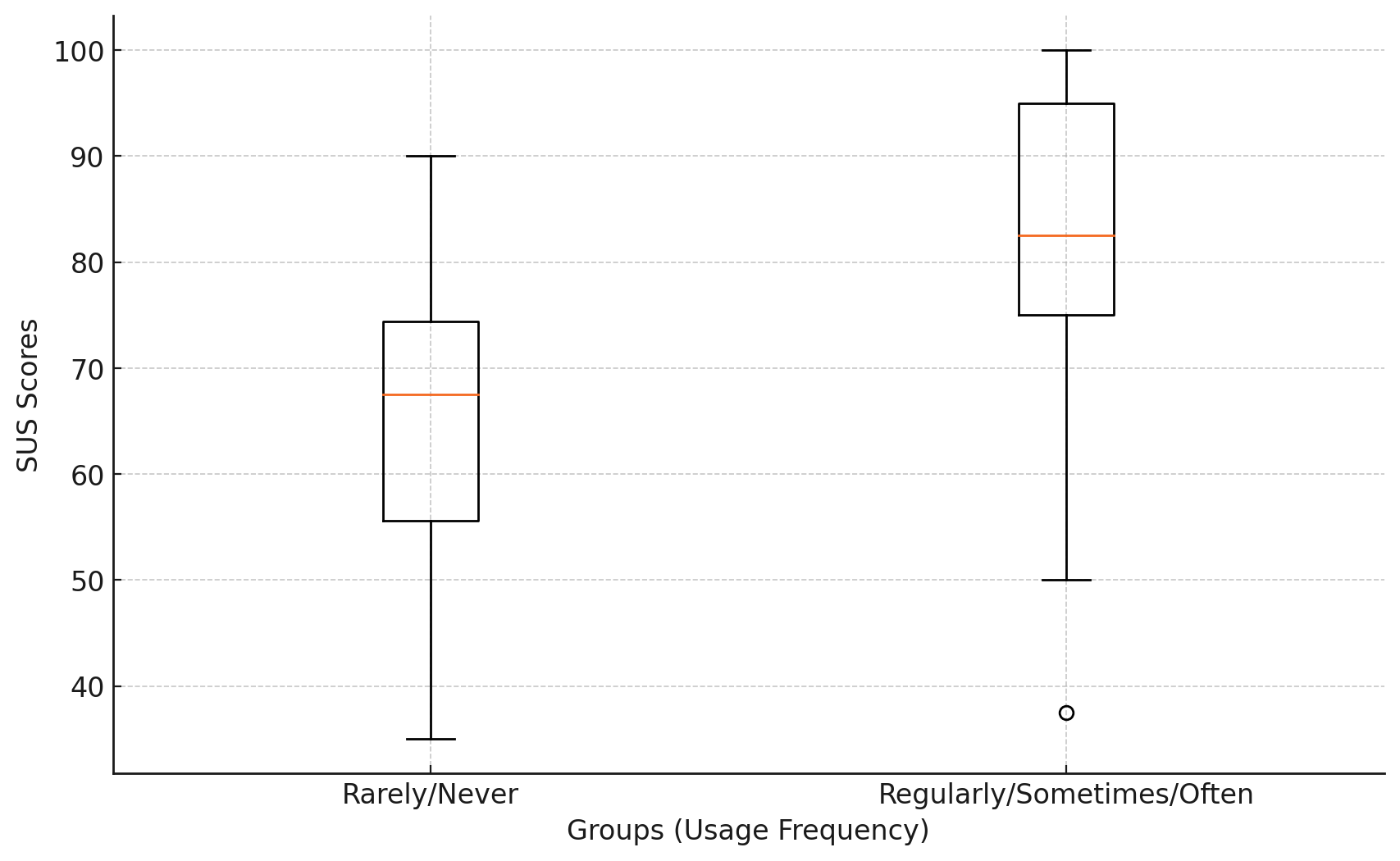}
    \caption{SUS score distribution by group (Rarely/Never vs. Regularly/Sometimes/Often).}
    \label{fig:boxplot_sus}
\end{figure}

The boxplot in Figure~\ref{fig:boxplot_sus} visually demonstrates the differences in score distributions between the two groups.  

The statistically significant difference in scores supports the hypothesis that frequency of use influences users' perceptions of system usability. Specifically, participants categorized as \textbf{Regularly/Sometimes/Often} tend to rate the system more favorably, achieving higher SUS scores compared to those categorized as \textbf{Rarely/Never}. These findings suggest that engagement frequency may play a crucial role in shaping user experiences and perceptions of usability.

In summary, the SUS score of \textbf{69.64} suggests that the system is largely usable with strengths in ease of use and user confidence. However, addressing areas related to complexity and support may further enhance user satisfaction and effectiveness. Additionally, the ANOVA analysis highlights the importance of engagement levels in influencing perceptions of usability. Future iterations of the system could benefit from design adjustments that streamline complex features, reduce the need for user support, and encourage more frequent use, ultimately aiming to improve the overall SUS score and user experience.

\section{Discussion}

Integrating generative AI for real-time 3D model generation revolutionizes various industries by making AR accessible and engaging.
Our approach allows users to create and interact with 3D content without specialized skills, fostering innovation.
We discuss diverse applications, enhanced user experiences, and the importance of user involvement in the model generation process.
We also discuss future research to improve the technology's capabilities and user-friendliness.

\subsection {System Requirements}

To ensure optimal performance for real-time 3D model generation in AR environments using our developed system, the recommended hardware specifications are as follows:
The recommended Graphics Processing Unit (GPU) is equipped with 16 GB of memory.
Additionally, the system is recommended to have a minimum total memory (RAM) of 32 GB.
The software requirements include an operating system compatible with Linux distributions for API development and Windows 10 or higher for AR environment development. 
The recommended AR headset is the Microsoft HoloLens 2 for sufficient display resolution, built-in sensors for environment mapping and object recognition, and support for interactive hand-tracking.

\subsection{Applications Across Industries}

The potential applications of our approach span multiple industries, including gaming, education, retail, and beyond. Gamers can use the system to generate and customize 3D objects for their favorite games, adding a new level of personalization and creativity to the gaming experience. This can also pave the way for user-generated content in gaming, with players sharing their custom 3D models within gaming communities. The ability to convert real-world objects into 3D models and visualize them in an AR environment is a game-changer for interior design. Designers can create, modify, and preview furniture and decor items in real-time, providing their clients with a realistic representation of their envisioned space. The system can transform online shopping by allowing customers to visualize products in 3D within their own environment. This not only enhances the shopping experience but also helps customers make more informed purchasing decisions.


\subsection{Democratization of 3D Model Creation}

Our research significantly democratizes 3D model creation, making it accessible to a wider audience. This fosters creativity and innovation, enabling individuals and small businesses to use AR without extensive technical expertise or resources. By empowering users to explore new ideas and applications, we break down barriers and broaden the scope of AR technologies, encouraging widespread adoption across various sectors.



\subsection{Future Research and Development}
Looking ahead, several areas for future research and development can enhance generative AI in AR environments. One key focus is improving object detection algorithms to handle complex and cluttered backgrounds, ensuring accurate isolation and conversion of objects. Exploring generative AI for dynamic and animated 3D models can create more engaging AR experiences, particularly for gaming and entertainment.

Enhancing scalability and efficiency is crucial for broader adoption. Optimizing computational requirements and processing times will allow deployment on a wider range of devices, including those with limited resources. AI models, as used for multilingual speech processing in secure communication systems ~\cite{Majid2024}, can similarly enhance AR by managing voice commands, improving object selection, and reducing manual zone selection. This would improve object selection accuracy and streamline the user experience, reducing the need for manual zone selection.

Finally, ongoing user evaluation and feedback are vital for continuous improvement. By conducting user studies and gathering insights from real-world applications, we can refine our approach to meet the evolving needs of AR users.

\section{Conclusion}
In conclusion, our research demonstrates the potential of generative AI to transform AR by enabling real-time 3D model generation from real-world objects and images. This enhances AR environments' immersive quality and interactivity. Advanced object detection and user-friendly zone selection address challenges like complex backgrounds and varying lighting, making 3D model generation more precise and accessible. The Shap-E model effectively creates detailed 3D representations, benefiting industries such as gaming, education, and retail by providing realistic, customizable 3D objects in real-time.

Our advancements contribute to AR and generative AI by addressing image-to-3D conversion challenges and promoting user involvement for accuracy and relevance. This technology has practical applications in retail and education, enhancing virtual product demonstrations and interactive learning. Our research pushes the boundaries of generative AI and AR, aiming for a seamless integration of physical and digital worlds for intuitive user experiences.

\bibliographystyle{plain}
\bibliography{BehravanHCII2025}

\begin{thebibliography}{10}

\bibitem{achlioptas2018learning}
Panos Achlioptas, Olga Diamanti, Ioannis Mitliagkas, and Leonidas Guibas.
\newblock Learning representations and generative models for 3{D} point clouds.
\newblock In {\em Proceedings of the International Conference on Machine Learning}, pages 40--49. PMLR, 2018.

\bibitem{BehravanVRST}
Majid Behravan and Denis Gračanin.
\newblock Generative multi-modal artificial intelligence for dynamic real-time context-aware content creation in augmented reality.
\newblock In {\em Proceedings of the 30th ACM Symposium on Virtual Reality Software and Technology}, VRST '24. Association for Computing Machinery, 2024.

\bibitem{behravan2024AIxVR}
Majid Behravan and Denis Gračanin.
\newblock Generative {AI} for context-aware 3{D} object creation using vision-language models in augmented reality.
\newblock In {\em Proceedings of the 7th IEEE International Conference on Artificial Intelligence \& eXtended and Virtual Reality}. IEEE, 2025.

\bibitem{Majid2024}
Majid Behravan, Elham Mohammadrezaei, Mohamed Azab, and Denis Gračanin.
\newblock Multilingual standalone trustworthy voice-based social network for disaster situations.
\newblock In {\em 2024 IEEE 15th Annual Ubiquitous Computing, Electronics \& Mobile Communication Conference (UEMCON)}, pages 264--270, 2024.

\bibitem{cao2020}
Michael~M. Bronstein, Joan Bruna, Taco Cohen, and Petar Veličković.
\newblock Geometric deep learning: Grids, groups, graphs, geodesics, and gauges.
\newblock {\em arXiv:2104.13478 [cs.LG]}, 2020.

\bibitem{chan2022efficient}
Eric~R. Chan, Connor~Z. Lin, Matthew~A. Chan, Koki Nagano, Boxiao Pan, Shalini~De Mello, Orazio Gallo, Leonidas Guibas, Jonathan Tremblay, Sameh Khamis, Tero Karras, and Gordon Wetzstein.
\newblock Efficient geometry-aware 3{D} generative adversarial networks.
\newblock In {\em Proceedings of the IEEE/CVF Conference on Computer Vision and Pattern Recognition}, pages 16123--16133, 2022.

\bibitem{chen2019learning}
Zhiqin Chen and Hao Zhang.
\newblock Learning implicit fields for generative shape modeling.
\newblock {\em Proceedings of the IEEE/CVF Conference on Computer Vision and Pattern Recognition}, pages 5939--5948, 2019.

\bibitem{Dasgupta-2020-a}
Archi Dasgupta, Mark Manuel, Rifat Mansur, Nabil Nowak, and Denis Gra{\v{c}}anin.
\newblock Towards real time object recognition for context awareness in mixed reality: A machine learning approach.
\newblock In {\em Proceedings of the 2020 IEEE Conference on Virtual Reality and 3D User Interfaces Abstracts and Workshops ({VRW})}, pages 262--268. IEEE, 22--26~ 2020.

\bibitem{gao2022get3d}
Jun Gao, Tianchang Shen, Zian Wang, Wenzheng Chen, Kangxue Yin, Daiqing Li, Or~Litany, Zan Gojcic, and Sanja Fidler.
\newblock {GET3D}: A generative model of high quality 3{D} textured shapes learned from images.
\newblock {\em Advances in Neural Information Processing Systems}, 35:31841--31854, 2022.

\bibitem{girshick2014rich}
Ross Girshick, Jeff Donahue, Trevor Darrell, and Jitendra Malik.
\newblock Rich feature hierarchies for accurate object detection and semantic segmentation.
\newblock In {\em Proceedings of the IEEE conference on computer vision and pattern recognition}, pages 580--587, 2014.

\bibitem{goodfellow2014generative}
Ian~J. Goodfellow, Jean Pouget-Abadie, Mehdi Mirza, Bing Xu, David Warde-Farley, Sherjil Ozair, Aaron Courville, and Yoshua Bengio.
\newblock Generative adversarial nets.
\newblock {\em Advances in neural information processing systems}, 27:2672--2680, 2014.

\bibitem{hanocka2019}
Rana Hanocka, Amir Hertz, Noa Fish, Raja Giryes, Shachar Fleishman, and Daniel Cohen-Or.
\newblock Mesh{CNN}: A network with an edge.
\newblock {\em ACM Transactions on Graphics}, 38(4):1--12, 2024.

\bibitem{hassan2022review}
Esraa Hassan, Nora El-Rashidy, and Fatma~M. Talaa.
\newblock Review: Mask r-cnn models.
\newblock {\em Nile Journal of Communication \& Computer Science}, 3(1):1--10, 2022.

\bibitem{he2017mask}
Kaiming He, Georgia Gkioxari, Piotr Dollár, and Ross Girshick.
\newblock Mask {R}-{CNN}.
\newblock arXiv:1703.06870 [cs.CV], 2017.

\bibitem{huang2023shapeclipper}
Zixuan Huang, Varun Jampani, Anh Thai, Yuanzhen Li, Stefan Stojanov, and James~M. Rehg.
\newblock Shape{C}lipper: Scalable 3{D} shape learning from single-view images via geometric and {CLIP}-based consistency.
\newblock In {\em Proceedings of the 2023 IEEE/CVF Conference on Computer Vision and Pattern Recognition (CVPR)}, pages 12912--12922, 2023.

\bibitem{hui2022neural}
Ka-Hei Hui, Ruihui Li, Jingyu Hu, and Chi-Wing Fu.
\newblock Neural wavelet-domain diffusion for 3{D} shape generation.
\newblock {\em ACM SIGGRAPH Asia}, 2022.

\bibitem{luo2021diffusion}
Shitong Luo and Wei Hu.
\newblock Diffusion probabilistic models for 3{D} point cloud generation.
\newblock {\em Proceedings of the IEEE/CVF Conference on Computer Vision and Pattern Recognition}, pages 2837--2845, 2021.

\bibitem{muller2011}
Roland~M. Mueller and Katja Thoring.
\newblock Understanding artifact knowledge in design science.
\newblock {\em Proceedings of the 17th Americas Conference on Information Systems}, 2011.

\bibitem{musyarofah2020mask}
M~Musyarofah, Valentina Schmidt, and Martin Kada2.
\newblock Object detection of aerial image using mask-region convolutional neural network (mask {R}-{CNN}).
\newblock {\em IOP Conference Series: Earth and Environmental Science}, 500:012090, 2020.

\bibitem{nash2022}
Charlie Nash, João Carreira, Jacob Walker, Iain Barr, Andrew Jaegle, Mateusz Malinowski, and Peter Battaglia.
\newblock Transframer: Arbitrary frame prediction with generative models.
\newblock arXiv:2203.09494 [cs.CV], 2022.

\bibitem{nobari2021}
Amin~Heyrani Nobari, Muhammad~Fathy Rashad, and Faez Ahmed.
\newblock Creative{GAN}: Editing generative adversarial networks for creative design synthesis.
\newblock {\em Proceedings of the ASME 2021 International Design Engineering Technical Conferences and Computers and Information in Engineering Conference}, page V03AT03A002, 2021.

\bibitem{redmon2016yolo}
Joseph Redmon, Santosh Divvala, Ross Girshick, and Ali Farhadi.
\newblock You only look once: Unified, real-time object detection.
\newblock In {\em Proceedings of the 2016 IEEE Conference on Computer Vision and Pattern Recognition (CVPR)}, pages 779--788, 2016.

\bibitem{redmon2018yolov3}
Joseph Redmon and Ali Farhadi.
\newblock {YOLO}v3: An incremental improvement.
\newblock arXiv:1804.02767 [cs.CV], 2018.

\bibitem{ren2015faster}
Shaoqing Ren, Kaiming He, Ross Girshick, and Jian Sun.
\newblock Faster {R}-{CNN}: Towards real-time object detection with region proposal networks.
\newblock {\em IEEE Transactions on Pattern Analysis and Machine Intelligence}, 39(6):1137--1149, 2017.

\bibitem{shue2023triplane}
J.~Ryan Shue, Eric~Ryan Chan, Ryan Po, Zachary Ankner, Jiajun Wu, and Gordon Wetzstein.
\newblock 3{D} neural field generation using triplane diffusion.
\newblock In {\em Proceedings of the IEEE/CVF Conference on Computer Vision and Pattern Recognition (CVPR)}, pages 20875--20886, 2023.

\bibitem{peize2020sparse}
Peize Sun, Rufeng Zhang, Yi~Jiang, Tao Kong, Chenfeng Xu, Masayoshi~Tomizuka Wei~Zhan, Lei Li, Changhu~Wang Zehuan~Yuan, and Ping Luo.
\newblock Sparse {R}-{CNN}: End-to-end object detection with learnable proposals.
\newblock arXiv:2011.12450 [cs.CV], 2020.

\bibitem{tibshirani1996regression}
Robert Tibshirani.
\newblock Regression shrinkage and selection via the {L}asso.
\newblock {\em Journal of the Royal Statistical Society: Series B (Methodological)}, 58(1):267--288, 1996.

\bibitem{yu2021pixelnerf}
Alex Yu, Vickie Ye, Matthew Tancik, and Angjoo Kanazawa.
\newblock Pixelnerf: Neural radiance fields from one or few images.
\newblock In {\em Proceedings of the 2020 IEEE/CVF Conference on Computer Vision and Pattern Recognition (CVPR)}, pages 5386--5396, 2021.

\bibitem{Yu-2016-a}
Lingyun Yu, Konstantinos Efstathiou, Petra Isenberg, and Tobias Isenberg.
\newblock {CAST}: Effective and efficient user interaction for context-aware selection in 3{D} particle clouds.
\newblock {\em IEEE Transactions on Visualization and Computer Graphics}, 22(1):886--895, January 2016.

\bibitem{zamir2018taskonomy}
Amir~R. Zamir, Alexander Sax, William Shen, Leonidas Guibas, Jitendra Malik, and Silvio Savarese.
\newblock Taskonomy: Disentangling task transfer learning.
\newblock In {\em Porceedinsg of the 2018 IEEE/CVF Conference on Computer Vision and Pattern Recognition}, pages 3712--3722, June 2018.

\bibitem{zeng2022latent}
Xiaohui Zeng, Arash Vahdat, Francis Williams, Zan Gojcic, Or~Litany, Sanja Fidler, and Karsten Kreis.
\newblock {LION}: Latent point diffusion models for 3{D} shape generation.
\newblock arXiv:2210.06978 [cs.CV], 2022.

\bibitem{zheng2023locally}
Xin-Yang Zheng, Hao Pan, Peng-Shuai Wang, Xin Tong, Yang Liu, and Heung-Yeung Shum.
\newblock Locally attentional {SDF} diffusion for controllable 3{D} shape generation.
\newblock arXiv:2305.04461 [cs.CV], 2023.

\bibitem{zhou2019objects}
Xingyi Zhou, Dequan Wang, and Philipp Krähenbühl.
\newblock Objects as points.
\newblock arXiv:1904.07850 [cs.CV], 2019.

\bibitem{fu2021auto}
Zhizhuo Zhou and Shubham Tulsiani.
\newblock Sparsefusion: Distilling view-conditioned diffusion for 3{D} reconstruction.
\newblock {\em CVPR}, 2023.

\bibitem{zhu2024lassonet}
Chen Zhu-Tian, Wei Zeng, Zhiguang Yang, Lingyun Yu, Chi-Wing Fu, and Huamin Qu.
\newblock Lasso{N}et: Deep {L}asso-selection of 3{D} point clouds.
\newblock {\em IEEE Transactions on Visualization and Computer Graphics}, 26(1):195--204, January 2020.

\end{thebibliography}

\end{document}